\documentclass[twocolumn,superscriptaddress,noshowpacs]{revtex4}
\usepackage{amsmath,amsthm,amssymb}
\usepackage{latexsym}
\usepackage{array}
\usepackage{amscd,graphicx}
\usepackage{bm,textcomp}
\usepackage[titletoc]{appendix}
\usepackage{epsfig}
\usepackage{epstopdf}
\usepackage{color}
\usepackage{mathrsfs}

\begin{document}

\title{Four-junction superconducting circuit}
\author{Yueyin Qiu}
\affiliation{Department of Physics, Fudan University, Shanghai 200433,China}
\affiliation{Beijing Computational Science Research Center, Beijing 100193, China}
\author{Wei Xiong}
\affiliation{Department of Physics, Fudan University, Shanghai 200433,China}
\affiliation{Beijing Computational Science Research Center, Beijing 100193, China}
\author{Xiao-Ling He}
\affiliation{School of Science, Zhejiang University of Science and Technology, Hangzhou,
Zhejiang 310023, China}
\author{Tie-Fu Li}
\affiliation{Institute of Microelectronics, Department of Microelectronics and Nanoelectronics and Tsinghua National Laboratory of Information Science and Technology, Tsinghua University, Beijing 100084, China}
\affiliation{Beijing Computational Science Research Center, Beijing 100193, China}
\author{J. Q. You}
\affiliation{Beijing Computational Science Research Center, Beijing 100193, China}
\date{\today }

\begin{abstract}
We develop a theory for the quantum circuit consisting of a superconducting
loop interrupted by four Josephson junctions and pierced by a magnetic flux
(either static or time-dependent). In addition to the similarity with the
typical three-junction flux qubit in the double-well regime, we demonstrate the difference of the four-junction circuit from its three-junction analogue, including its
advantages over the latter. Moreover, the four-junction circuit in
the single-well regime is also investigated. Our theory provides a tool to explore
the physical properties of this four-junction superconducting circuit.
\end{abstract}

\maketitle



Superconducting quantum circuits based on Josephson junctions exhibit
macroscopic quantum coherence and can be used as qubits for quantum
information processing (see, e.g., Refs.~\cite%
{nakamura-99,bouchiat-98new,vion-02,Chiorescu-03,yu-02,martinis-02,chiorescu-04,wallraff-04,yan-15}%
). Behaving as artificial atoms, these circuits can also be utilized to
demonstrate novel atomic-physics and quantum-optics phenomena, including
those that are difficult to observe or even do not occur in natural atomic
systems~\cite{you-11}. As a rough distinction, there are three types of
superconducting qubits, i.e., charge~\cite{nakamura-99,bouchiat-98new},
flux~\cite{Chiorescu-03,Mooij-99} and phase qubits~\cite%
{yu-02,martinis-02,martinis-09new}. In the charge qubit, where the charge
degree of freedom dominates, two discrete Cooper-pair states are coupled via
a Josephson coupling energy~\cite{nakamura-99,bouchiat-98new}. In contrast,
the phase degree of freedom dominates in both flux~\cite{Mooij-99} and phase qubits~\cite{yu-02,martinis-02}.

The typical flux qubit is composed of a superconducting loop interrupted by
three Josephson junctions~\cite{Mooij-99}. Similar to other types of
superconducting qubits, it exhibits good quantum coherence and can be tuned
externally. Recent experimental measurements~\cite{yan-15} showed that the decoherence
time of the three-junction flux qubit can be longer than 40~$\mu$s.
Due to the convenience in sample fabrication (i.e., the
double-layer structure fabrication by the shadow evaporation technique~\cite%
{Kuzmin-91new}), a superconducting loop interrupted by four Josephson
junctions was also used as the flux qubit. The experiments~\cite%
{Bertet-05} showed that this four-junction
flux qubit behaves similar to the three-junction flux qubit.
Also, two four-junction flux qubits were interacting experimentally via a coupler~\cite{Niskanen-07new}, similar to the interqubit coupling mediated by a high-excitation-energy quantum object~\cite{Ashhab}.
The theory of
the three-junction flux circuit with a static flux bias was well developed~%
\cite{Orlando-99}, but a theory for the four-junction circuit lacks
because adding one Josephson junction more to the superconducting loop makes
the problem more complex.

In this paper, we develop a theory for the
four-junction circuit with either a static or time-dependent flux bias.
In addition to the similarity with the three-junction circuit, we
demonstrate the difference from the three-junction circuit due to the different sizes of the two smaller Josephson junctions in the four-junction circuit. We find that the four-junction circuit with only one smaller junction has a broader parameter range to achieve a flux qubit in the double-well regime than the three-junction circuit.
Moreover, for the four-junction circuit with two identical smaller junctions, the circuit can be used as a qubit better than the three-junction circuit, because it becomes more robust against the state leakage from the qubit subspace to the third level. This can be a useful advantage of the four-junction circuit over the three-junction circuit when used as a qubit. Also, we study the four-junction circuit in the single-well regime,
which was not exploited before. Our theory can provide a
useful tool to explore the physical properties of this four-junction
superconducting circuit.

\vspace{1cm} \noindent \textbf{\large Results}

\begin{figure}[tbp]
\includegraphics[width=8.8cm]{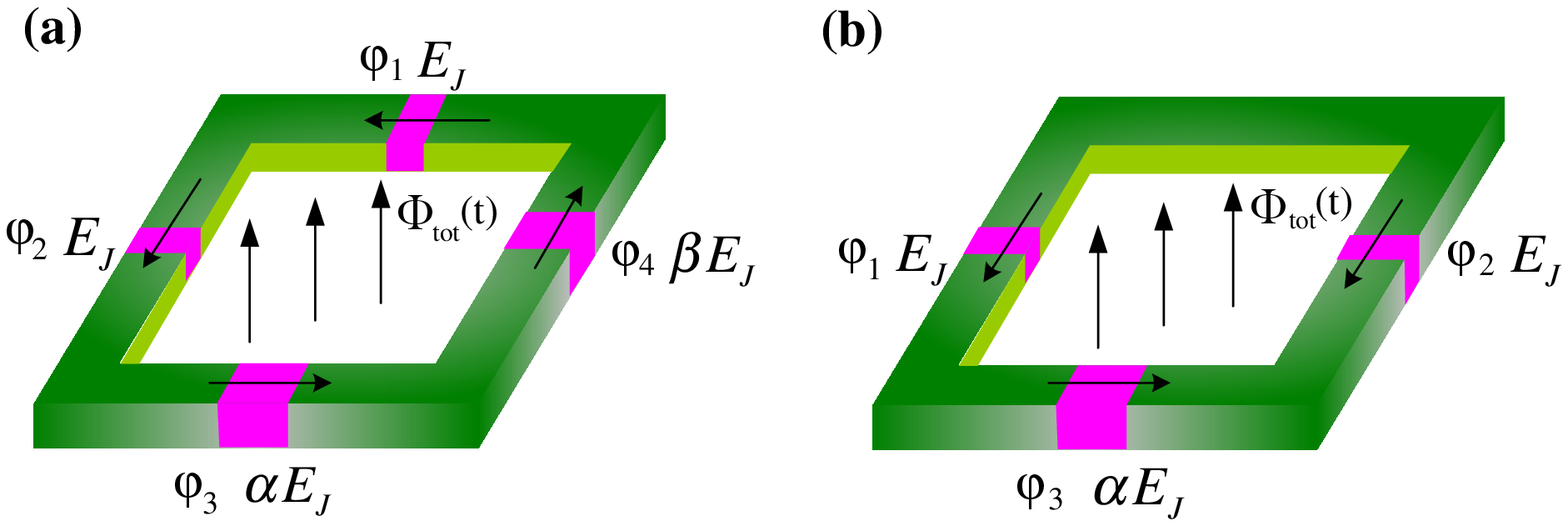}
\caption{(color online) Schematic diagram of the considered superconducting
circuits. (a)~Superconducting loop interrupted by four Josephson junctions
and pierced by a total magnetic flux, $\Phi_{\mathrm{tot}}(t)$, which
includes the externally applied flux and the inductance-induced flux. Here
two of the four junctions have identical Josephson coupling energy $E_{J}$
and capacitance $C$. Among other two junctions, one has Josephson coupling
energy $\protect\alpha E_{J}$ and capacitance $\protect\alpha C$, and the
other has Josephson coupling energy $\protect\beta E_{J}$ and capacitance $%
\protect\beta C$, with $0<\protect\alpha,\protect\beta<1$.
(b)~Superconducting loop interrupted by three Josephson junctions and
pierced by a total magnetic flux $\Phi_{\mathrm{tot}}(t)$, where two
junctions have identical Josephson coupling energy $E_{J}$ and capacitance $%
C $, while the third one has Josephson coupling energy $\protect\alpha E_{J}$
and capacitance $\protect\alpha C$, with $0<\protect\alpha<1$. In both (a)
and (b), each red component denotes the thin insulator layer of a Josephson
junction, and an arrow along the loop denotes the assigned direction of the
phase drop across the corresponding Josephson junction. Note that each phase
drop can be chosen along either the clockwise or counter-clockwise
direction, but once the direction is fixed, the phase drop is positive along
it.}
\label{system}
\end{figure}

\vspace{.1cm}\noindent \textbf{Four-junction superconducting circuit.}~~~(1)~%
\textit{The total Hamiltonian.} Let us consider a superconducting loop
interrupted by four Josephson junctions and pierced by a magnetic flux [see
Fig.~\ref{system}(a)], where the first and second junctions have identical
Josephson coupling energy $E_{J}$ and capacitance $C$ (i.e., $E_{Ji}=E_{J}$
and $C_{i}=C$, with $i=1,2$), while the third and fourth junctions are
reduced as $E_{J3}=\alpha E_{J}$, $E_{J4}=\beta E_{J}$, $C_{3}=\alpha C$,
and $C_{4}=\beta C$, with $0<\alpha,\beta<1$. The phase drops $\varphi _{i}$
($i=1,2,3,4$) through these four Josephson junctions are constrained by the
fluxoid quantization
\begin{equation}
\varphi _{1}+\varphi _{2}+\varphi _{3}+\varphi _{4}+2\pi f_{\mathrm{tot}%
}(t)=0,  \label{eq1}
\end{equation}
where $f_{\mathrm{tot}}(t)=\Phi _{\mathrm{tot}}(t)/\Phi _{0}$, with $\Phi _{%
\mathrm{tot}}(t)$ being the total magnetic flux in the loop (which includes
the externally applied flux, either static or time-dependent, and the
inductance-induced flux owing to the persistent current in the loop) and $%
\Phi_{0}=h/2e$ being the flux quantum.

The kinetic energy of the four-junction circuit is the electrostatic energy~%
\cite{Makhlin-01} stored in the junction capacitors, which can be written as
\begin{equation}
\mathcal{T}=\frac{1}{2}\underset{i=1}{\overset{4}{\sum }}C_{i}V_{i}^{2},
\end{equation}%
where $V_{i}=(\Phi _{0}/2\pi )\dot{\varphi}_{i}$ is the voltage across the $%
i $th junction. Using the the fluxoid quantization condition in Eq.~(\ref%
{eq1}), we can rewrite the kinetic energy as
\begin{eqnarray}
\mathcal{T}\! &\!=\!&\!\frac{C}{2}\left( \frac{\Phi _{0}}{2\pi }\right) ^{2}%
\bigg\{\dot{\varphi}_{1}^{2}+\dot{\varphi}_{2}^{2}+\alpha \dot{\varphi}%
_{3}^{2} \\
&&\!+\beta \left[ \dot{\varphi}_{1}+\dot{\varphi}_{2}+\dot{\varphi}_{3}+2\pi
\dot{f}_{\mathrm{tot}}\right]^{2} \bigg\}.  \notag
\end{eqnarray}%
We introduce a phase transformation
\begin{eqnarray}
&&\varphi _{1}=\frac{\varphi }{\sqrt{2}}+b_{+}\varphi _{+}+b_{-}\varphi
_{-}+\alpha b\xi ,  \notag \\
&&\varphi _{2}=-\frac{\varphi }{\sqrt{2}}+b_{+}\varphi _{+}+b_{-}\varphi
_{-}+\alpha b\xi ,  \label{PT} \\
&&\varphi _{3}=-\frac{2\beta b_{+}}{\alpha +\beta -\lambda _{+}}\varphi _{+}-%
\frac{2\beta b_{-}}{\alpha +\beta -\lambda _{-}}\varphi _{-}+b\xi ,  \notag
\end{eqnarray}%
where
\begin{eqnarray}
\xi \! &\!=\!&\!f_{\mathrm{tot}}-f_{e},~~~~b=-\frac{2\pi \beta }{\alpha
+\beta +2\alpha \beta },  \notag \\
b_{\pm }\! &\!=\!&\!\frac{|\alpha +\beta -\lambda _{\pm }|}{\sqrt{2(\alpha
+\beta )^{2}+4\beta ^{2}-4(\beta +\alpha )\lambda _{\pm }+2\lambda _{\pm
}^{2}}}, \\
\lambda _{\pm }\! &\!=\!&\!\frac{(1+\alpha +3\beta )\pm \sqrt{1+(\alpha
-\beta )^{2}+8\beta ^{2}+2\left( \beta -\alpha \right) }}{2},  \notag
\end{eqnarray}%
with $f_{e}=\Phi _{e}/\Phi _{0}$ being the reduced static magnetic flux
applied to the superconducting loop. The electrostatic energy $\mathcal{T}$
can then be converted to a quadratic form
\begin{equation}
\mathcal{T}=\frac{C}{2}\left( \frac{\Phi _{0}}{2\pi }\right) ^{2}(\dot{%
\varphi}^{2}+\Gamma _{+}\dot{\varphi}_{+}^{2}+\Gamma _{-}\dot{\varphi}%
_{-}^{2}+\Gamma _{\xi }\dot{\xi}^{2}),
\end{equation}%
where
\begin{eqnarray}
\Gamma _{\pm }\!\! &=&\!\!2b_{\pm }^{2}\left[ 1+2\beta -\frac{4\beta ^{2}}{%
\alpha +\beta -\lambda _{\pm }}+\frac{2\beta ^{2}(\alpha +\beta )}{(\alpha
+\beta -\lambda _{\pm })^{2}}\right] ,  \notag \\
\Gamma _{\xi }\! &\!=\!&\!\left( 2\alpha ^{2}+4\beta \alpha ^{2}+\alpha
+\beta +4\alpha \beta \right) b^{2} \\
&&+4\pi \beta (1+2\alpha )b+4\pi ^{2}\beta .  \notag
\end{eqnarray}

The total Josephson coupling energy of the four-junction circuit is
\begin{eqnarray}
U\! &\!=\!&\!\underset{i=1}{\overset{4}{\sum }}E_{Ji}(1-\cos \varphi _{i})
\notag \\
\! &\!=\!&\!E_{J}[2+\alpha +\beta -\cos \varphi _{1}-\cos \varphi _{2}
\notag \\
&&\!-\alpha \cos \varphi _{3}-\beta \cos (\varphi _{1}+\varphi _{2}+\varphi
_{3}+2\pi f_{\mathrm{tot}})]  \notag \\
\! &\!=\!&\!E_{J}\bigg[2+\alpha +\beta -\cos \left( \frac{\varphi }{\sqrt{2}}%
+b_{+}\varphi _{+}+b_{-}\varphi _{-}+\alpha b\xi \right)  \notag \\
&&\!-\cos \left( -\frac{\varphi }{\sqrt{2}}+b_{+}\varphi _{+}+b_{-}\varphi
_{-}+\alpha b\xi \right)  \notag \\
&&\!-\alpha \cos \left( -\frac{2\beta b_{+}}{\alpha +\beta -\lambda _{+}}%
\varphi _{+}-\frac{2\beta b_{-}}{\alpha +\beta -\lambda _{-}}\varphi
_{-}+b\xi \right)  \notag \\
&&\!-\beta \cos \bigg(\frac{2\left( \alpha -\lambda _{+}\right) b_{+}}{%
\alpha +\beta -\lambda _{+}}\varphi _{+}+\frac{2\left( \alpha -\lambda
_{-}\right) b_{-}}{\alpha +\beta -\lambda _{-}}\varphi _{-}  \notag \\
&&~~~~~~~~~~+\left( 2\alpha b+b+2\pi \right) \xi +2\pi f_{e}\bigg)\bigg].
\end{eqnarray}%
Also, there is the inductive energy due to the inductance $L$ of the
superconducting loop~\cite{you-05new}:
\begin{equation}
U_{L}=\frac{\Phi _{0}^{2}}{2L}(f_{\mathrm{tot}}-f_{\mathrm{ext}})^{2},
\end{equation}%
where the reduced externally-applied magnetic flux $f_{\mathrm{ext}}$ can
generally be written as a sum of the static and time-dependent fluxes, i.e.,
$f_{\mathrm{ext}}=f_{e}+f_{a}(t)$, with $f_{a}(t)\equiv \Phi _{a}(t)/\Phi
_{0}$ being the reduced time-dependent magnetic field applied to the
four-junction loop. When including this inductive energy, the total
potential energy of the four-junction circuit is written as
\begin{equation}
\mathcal{U}=U(\varphi ,\varphi _{+},\varphi _{-},\xi )+\frac{\Phi _{0}^{2}}{%
2L}(\xi -f_{\mathrm{a}})^{2}.
\end{equation}

The Lagrangian of the four-junction circuit is
\begin{eqnarray}
\mathcal{L}\! &\!=\!&\!\mathcal{T}-\mathcal{U}  \notag \\
&=&\frac{C}{2}\left( \frac{\Phi _{0}}{2\pi }\right) ^{2}(\dot{\varphi}%
^{2}+\Gamma _{+}\dot{\varphi}_{+}^{2}+\Gamma _{-}\dot{\varphi}%
_{-}^{2}+\Gamma _{\xi }\dot{\xi}^{2})  \notag \\
&&-U(\varphi ,\varphi _{+},\varphi _{-},\xi )-\frac{\Phi _{0}^{2}}{2L}(\xi
-f_{\mathrm{a}})^{2}.
\end{eqnarray}%
where we assign $\varphi $, $\varphi _{\pm }$, and $\xi $ as the canonical
coordinates. The corresponding canonical momenta $P=\partial \mathcal{L}%
/\partial \dot{\varphi}$, $P_{\pm }=\partial \mathcal{L}/\partial \dot{%
\varphi}_{\pm }$, and $P_{\xi }=\partial \mathcal{L}/\partial \dot{\xi}$ are
\begin{eqnarray}
P\! &\!=\!&\!C\left( \frac{\Phi _{0}}{2\pi }\right) ^{2}\dot{\varphi},
\notag \\
P_{\pm }\! &\!=\!&\!C\left( \frac{\Phi _{0}}{2\pi }\right) ^{2}\Gamma _{\pm }%
\dot{\varphi}_{\pm }, \\
P_{\xi }\! &\!=\!&\!C\left( \frac{\Phi _{0}}{2\pi }\right) ^{2}\Gamma _{\xi }%
\dot{\xi}.  \notag
\end{eqnarray}%
Therefore, the Hamiltonian of the four-junction circuit is given by
\begin{eqnarray}
H\! &\!=\!&\!\sum_{i}P_{i}\dot{\varphi}_{i}-\mathcal{L}  \notag \\
&=&4E_{C}\left( P^{2}+\frac{P_{+}^{2}}{\Gamma _{+}}+\frac{P_{-}^{2}}{\Gamma
_{-}}+\frac{P_{\xi }^{2}}{\Gamma _{\xi }}\right)  \notag \\
&&\!+U(\varphi ,\varphi _{+},\varphi _{-},\xi )+\frac{\Phi _{0}^{2}}{2L}(\xi
-f_{\mathrm{a}})^{2},
\end{eqnarray}%
where $E_{C}=e^{2}/(2C)$ is the single-particle charging energy of the
Josephson junction. In comparison with the previous work in Ref.~\cite%
{Orlando-99} for the three-junctions flux qubit, a new degree of freedom $%
\xi $ is included in the Hamiltonian, so that the Hamiltonian can also apply
to the case when the superconducting loop contains a time-dependent magnetic flux.

\vspace{0.3cm} (2)~\textit{The reduced Hamiltonian.} The total Hamiltonian
of the four-junction circuit can be rewritten as
\begin{equation}
H=4E_{C}\left( P^{2}+\frac{P_{+}^{2}}{\Gamma _{+}}+\frac{P_{-}^{2}}{\Gamma
_{-}}\right) +U(\varphi ,\varphi _{+},\varphi _{-},\xi )+H_{\mathrm{osc}},~
\end{equation}%
where
\begin{equation}
H_{\mathrm{osc}}=\frac{4E_{C}}{\Gamma _{\xi }}P_{\xi }^{2}+\frac{\Phi
_{0}^{2}}{2L}(\xi -f_{\mathrm{a}})^{2},  \label{osc4}
\end{equation}%
Quantum mechanically, the canonical momenta can be written as $P=-i\hbar
\partial /\partial \varphi $, $P_{\pm }=-i\hbar \partial /\partial \varphi
_{\pm }$, and $P_{\xi }=-i\hbar \partial /\partial \varphi _{\xi }$ in the
canonical-coordinate representation.

Note that the Hamiltonian $H_{\mathrm{osc}}$ in Eq.~(\ref{osc4}) can be
rewritten as
\begin{equation}
H_{\mathrm{osc}}=\frac{4E_{C}}{\Gamma _{\xi }}P_{\xi }^{2}+\frac{\Phi
_{0}^{2}}{2L}\xi ^{2}-\frac{\Phi _{0}^{2}}{L}\xi f_{\mathrm{a}}(t),
\end{equation}%
i.e., a harmonic oscillator driven by a time-dependent magnetic flux $f_{%
\mathrm{a}}(t)$. The angular frequency of this harmonic oscillator is
\begin{equation}
\omega _{\mathrm{osc}}=\frac{1}{\sqrt{\Gamma _{\xi }CL}}.
\end{equation}%
With the parameters achieved in experiments for the flux qubit~\cite%
{Bertet-05,Zant-94}, $\alpha \sim 0.7$, $C\sim 8~\mathrm{fF}$, and $L\sim 10~%
\mathrm{pH}$. Moreover, $\beta \sim \alpha $, so $\omega _{\mathrm{osc}%
}/2\pi \sim 1\times 10^{3}$~GHz. For the four-junction flux qubit, the
energy gap $\Delta $ between the lowest two levels is typically $\Delta \sim
1$-$10$~GHz~\cite{Bertet-05,Niskanen-07new}, which is much
smaller than $\omega _{\mathrm{osc}}/2\pi \sim 1\times 10^{3}$~GHz. Usually,
the time-dependent magnetic flux $f_{\mathrm{a}}(t)$ applied to the
four-junction loop is a microwave wave with $\omega _{a}/2\pi \sim 1$-$10$%
~GHz, which is also much smaller than $\omega _{\mathrm{osc}}/2\pi $.
Because $\Delta \ll \omega _{\mathrm{osc}}/2\pi $ and the flux $f_{\mathrm{a}%
}(t)$ is also very off resonance from the harmonic oscillator (i.e., $\omega
_{a}\ll \omega _{\mathrm{osc}}$), the oscillator is nearly kept in the
ground state at a low temperature. Then, using the adiabatic approximation
to eliminate the degree of freedom of the oscillator, the Hamiltonian of the
four-junction circuit can be reduced to
\begin{equation}
H=4E_{C}\left( P^{2}+\frac{P_{+}^{2}}{\Gamma _{+}}+\frac{P_{-}^{2}}{\Gamma
_{-}}\right) +U(\varphi ,\varphi _{+},\varphi _{-},\xi ).  \label{Hfull}
\end{equation}%
Also, both $L$ and the persistent current $I$ of the superconducting loop
are small, so that $IL/\Phi _{0}\sim 10^{-3}$~\cite{Orlando-99}. This
inductance-induced flux is much smaller than the externally applied magnetic
flux $f_{\mathrm{ext}}=f_{e}+f_{a}(t)$. Therefore, the total flux $f_{%
\mathrm{tot}}$ can also be approximately written as $f_{\mathrm{tot}}\simeq
f_{e}+f_{a}(t)$.

Below we first study the static-flux case, i.e., only a static magnetic flux
is applied to the four-junction loop. In this case, $f_{\mathrm{tot}}\simeq
f_{e}$, so $\xi \simeq 0$. The phase transformation in Eq.~(\ref{PT})
becomes
\begin{eqnarray}
&&\varphi _{1}=\frac{\varphi }{\sqrt{2}}+b_{+}\varphi _{+}+b_{-}\varphi _{-},
\notag \\
&&\varphi _{2}=-\frac{\varphi }{\sqrt{2}}+b_{+}\varphi _{+}+b_{-}\varphi
_{-},  \label{19} \\
&&\varphi _{3}=-\frac{2\beta b_{+}}{\alpha +\beta -\lambda _{+}}\varphi _{+}-%
\frac{2\beta b_{-}}{\alpha +\beta -\lambda _{-}}\varphi _{-},  \notag
\end{eqnarray}%
and the Hamiltonian of the four-junction circuit in Eq.~(\ref{Hfull}) is
further reduced to
\begin{equation}
H_{0}=4E_{C}\left( P^{2}+\frac{P_{+}^{2}}{\Gamma _{+}}+\frac{P_{-}^{2}}{%
\Gamma _{-}}\right) +U(\varphi ,\varphi _{+},\varphi _{-}),  \label{H0}
\end{equation}%
with $U(\varphi ,\varphi _{+},\varphi _{-})\equiv U(\varphi ,\varphi
_{+},\varphi _{-},\xi )|_{\xi =0}$.

\begin{figure*}[tbp]
\includegraphics[scale=0.75]{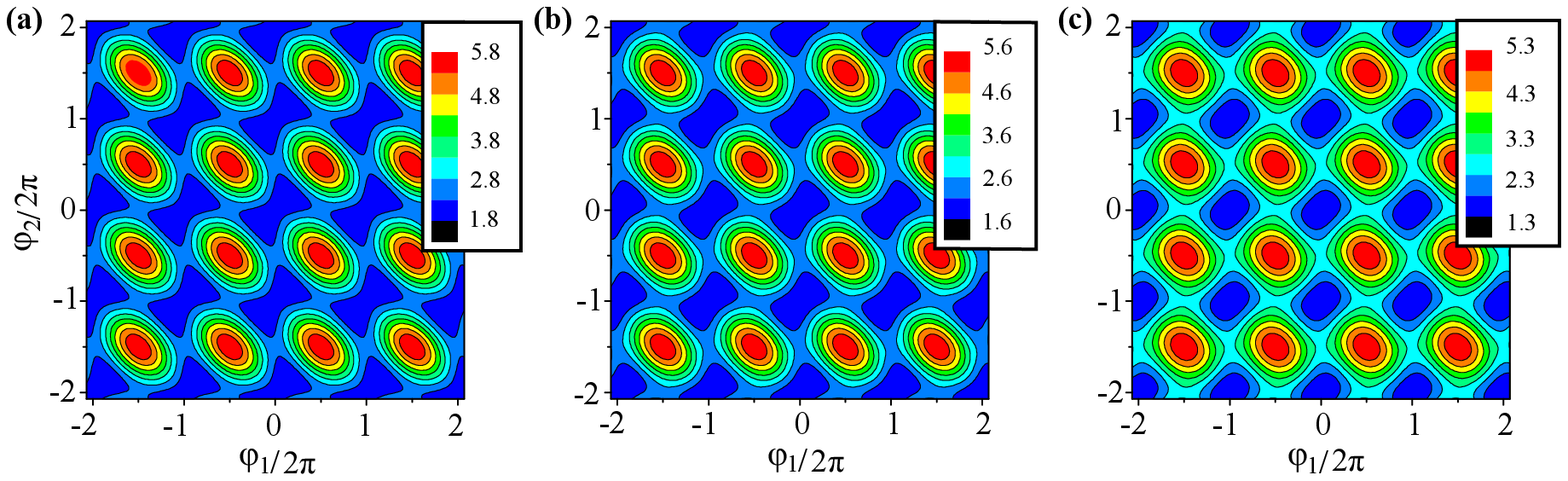}
\caption{(color online) Contour plots of the potential $U(\protect\varphi_1,%
\protect\varphi_2,\protect\varphi_3)$ at $\protect\varphi_3=0$ and $f_e=1/2$%
, where (a)~$\protect\alpha=1$, $\protect\beta=0.8$, (b)~$\protect\alpha=1$,
$\protect\beta=0.6$, and (c)~$\protect\alpha=1$, $\protect\beta=0.3$.}
\label{potential}
\end{figure*}

Figure~\ref{potential} shows the contour plots of the potential $U(\varphi
,\varphi _{+},\varphi _{-})\equiv U(\varphi _{1},\varphi _{2},\varphi _{3})$
in the two-dimensional subspace spanned by $\varphi _{1}$ and $\varphi _{2}$
for $f_{e}=1/2$, where $\varphi _{i}$ ($i=1,2,3$) are related to $\varphi $
and $\varphi _{\pm }$ by Eq.~(\ref{19}). For a three-junction flux qubit, $%
\alpha $ is usually in the range of $1/2<\alpha <1$. When $0<\alpha <1/2$, each
double well in the potential is reduced to a single well~\cite{Orlando-99},
so the flux qubit in the double-well regime is converted to a flux qubit in the single-well regime. For the
four-junction circuit, there are wider ranges of parameters to achieve a
flux qubit. For instance, in the case of three identical Josephson junctions
(i.e., $\alpha =1$ and $0<\beta <1$), when $\beta >1/3$, the potential $%
U(\varphi _{1},\varphi _{2},\varphi _{3})$ has two energy minima in the unit
cell of three-dimensional periodic lattice at $\varphi _{1}=\varphi
_{2}=\varphi _{3}=\pm \varphi ^{\ast }$ mod $2\pi $, where
\begin{equation}
\varphi ^{\ast }=\arcsin \left( \sqrt{\frac{3\beta -1}{4\beta }}\right) .
\end{equation}%
A flux qubit in the double-well potential can then be achieved
in the parameter range of $1/3<\beta <1$, which is broader than the range of
$1/2<\alpha <1$ for the three-junction flux qubit.
Figures~\ref{potential}(a) and \ref{potential}(b) show a section of $%
U(\varphi _{1},\varphi _{2},\varphi _{3})$ at $\varphi _{3}=0$.
Corresponding to the above-mentioned two minima, a figure-eight-shaped
double well exists in each unit cell of the periodic lattice in the
two-dimensional subspace. When $\beta <1/3$, each figure-eight-shaped double
well in the $\varphi _{3}=0$ section of the potential is reduced to a single
well [see Fig.~\ref{potential}(c)], with only one minimum in the unit cell
at $\varphi _{1}=\varphi _{2}=0$ mod $2\pi $. This corresponds to a flux qubit in the single-well regime achieved in the four-junction superconducting circuit.

\begin{figure*}[tbp]
\includegraphics[scale=0.9]{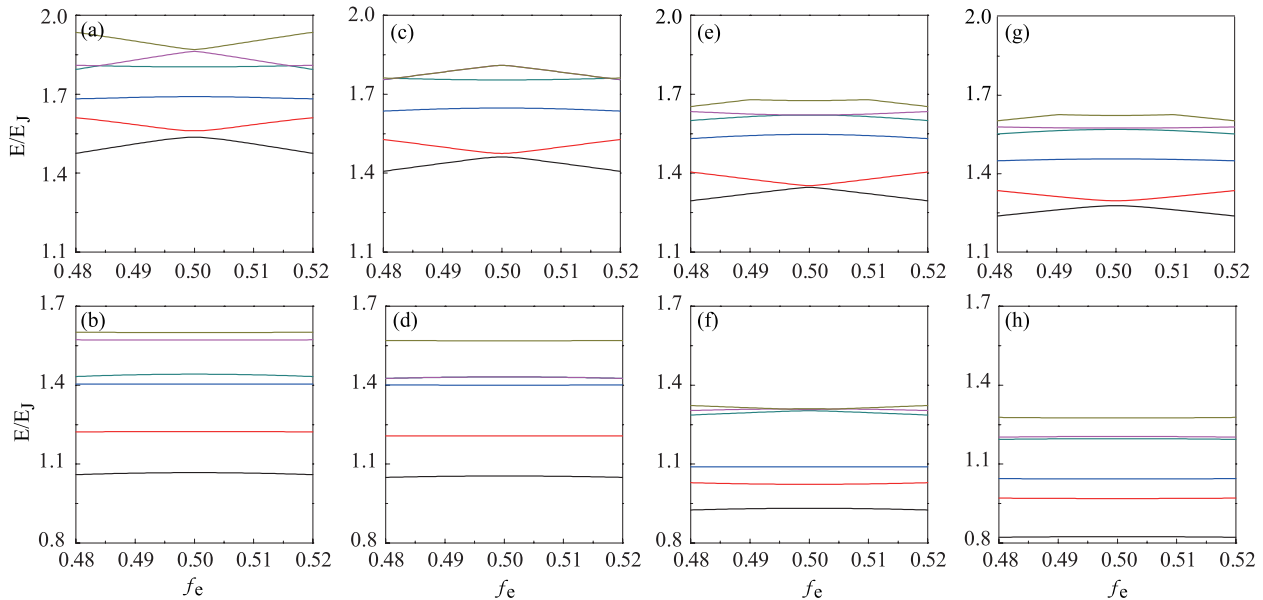}
\caption{(color online) Energy spectra of the superconducting circuits
versus the reduced static flux $f_{e}$. (a)~$\protect\alpha =0.7$ and (b)~$%
0.4$ in the case of three-junction circuit; (c)~$\protect\alpha =1$ and $%
\protect\beta =0.6$, (d)~$\protect\alpha =1$ and $\protect\beta =0.3$, (e)~$%
\protect\alpha=\protect\beta =0.6$, (f)~$\protect\alpha =\protect\beta %
=0.3$, (g)~$\protect\alpha =0.5$ and $\protect\beta %
=0.6$, and (h)~$\protect\alpha =0.2$ and $\protect\beta=0.3$ in the case of four-junction circuit. In this figure and the following one, we choose $E_{J}=50E_{C}$}.
\label{energy_50}
\end{figure*}

\vspace{.3cm}\noindent \textbf{Energy spectrum.} The energy spectrum and
eigenstates of the four-junction circuit are determined by
\begin{equation}
H_{0}\Psi (\boldsymbol{\varphi})=E\Psi (\boldsymbol{\varphi }),
\label{eigen}
\end{equation}
where $\boldsymbol{\varphi}\equiv (\varphi,\varphi_{+},\varphi_{-})
=(\varphi_1,\varphi_2,\varphi_3)$ is a three-dimensional vector in the phase
space. Equation~(\ref{eigen}) is just like the quantum mechanical problem of
a particle moving in a three-dimensional periodic potential $U(\boldsymbol{%
\varphi })$. Thus, the solution of it has the Bloch-wave form
\begin{equation}
\Psi (\boldsymbol{\varphi })=e^{i\boldsymbol{k}\cdot\boldsymbol{\varphi}}u(%
\boldsymbol{\varphi}),
\end{equation}
where $\boldsymbol{k}$ is a wavevector and $u(\boldsymbol{\varphi})$ is a
periodic function in the phases of $\varphi_i$ ($i=1,2,3$). Also, $\Psi (%
\boldsymbol{\varphi })$ should be periodic in the phases of $\varphi_i$. To
ensure this, the wavefunction $\Psi (\boldsymbol{\varphi })$ is constrained
by $\boldsymbol{k}=0$. Then, $\Psi (\boldsymbol{\varphi })$ can be written
as
\begin{equation}
\Psi (\boldsymbol{\varphi })=u(\boldsymbol{\varphi}) =\sum_{\boldsymbol{K}%
}a_{\boldsymbol{K}}e^{i\boldsymbol{K}\cdot\boldsymbol{\varphi}},
\label{wavefunction}
\end{equation}
where $\boldsymbol{K}$ is a reciprocal lattice vector. Substituting Eq.~(\ref%
{wavefunction}) into Eq.~(\ref{eigen}), we then obtain an equation similar to the central equation in the theory of energy bands~\cite{Kittel}. Numerically solving this equation, we can obtain the
energy spectrum and eigenstates of the Hamiltonian $H_0$.

For the three-junction flux qubit, an approximate tight-binding solution was obtained in Ref.~\cite{Orlando-99} by projecting the Schr\"{o}dinger equation onto the qubit subspace, where the needed tunneling matrix elements were estimated using the WKB method. For the four-junction case, such an approximate tight-binding solution can also be derived, but it is difficult to calculate the tunneling matrix elements via the WKB method, because a three-dimensional potential is involved in the four-junction circuit. Thus, we resort to the numerical approach to solve the Schr\"{o}dinger equation in Eq.~(\ref{eigen}). With this numerical approach, we can obtain the results for both the flux qubit and the three-level system.

Figure~\ref{energy_50} shows the energy levels of the four-junction circuit
versus the reduced static flux $f_{e}$, in comparison with the
three-junction circuit. In the case of four-junction circuit, when the
lowest two or three levels are considered, the energy spectrum with $\alpha
=1$ and $\beta =0.6$ is similar to the energy spectrum with $\alpha =0.7$ in
the case of three-junction circuit [comparing Fig.~\ref{energy_50}(c) with
Fig.~\ref{energy_50}(a)]. Because the lowest two levels are well separated
from other levels, both three- and four-junction circuits can be utilized as
quantum two-level systems (i.e., flux qubits). In this case, the flux qubit
can be modeled as
\begin{equation}
H_{0}=\frac{1}{2}\left( \varepsilon \sigma _{z}+\Delta \sigma _{x}\right) ,
\end{equation}%
where the tunneling amplitude $\Delta $ corresponds to the energy difference
between the two lowest-energy levels at $f_{e}=1/2$, and $\varepsilon
=2I_{p}\Phi _{0}(f_{e}-1/2)$ is the bias energy due to the external flux,
with $I_{p}$ being the maximal persistent current circulating in the loop.
Here the maximal persistent current $I_{p}$ can be approximately calculated
as~\cite{Orlando-99} $I_{p}\approx |\Phi _{0}^{-1}\partial E_{0}/\partial
f_{e}|$ at a value of $f_{e}$ considerably away from $f_{e}=1/2$, where $%
E_{0}$ is the energy level of the ground state of the system. The Pauli
operators $\sigma _{z}$ and $\sigma _{x}$ are represented using the two
(i.e., the clockwise and counter-clockwise) persistent-current states.
Moreover, similar to the three-junction circuit, the four-junction circuit
can also be used as a quantum three-level system (qutrit) owing to the considerable
separation of the third energy level from other higher levels as well. When
reducing the smallest junction to, e.g., $\beta =0.3$ in the four-junction
circuit [see Fig.~\ref{energy_50}(d)], only the lowest two levels are well
separated from other levels, similar to the case of three-junction circuit
in Fig.~\ref{energy_50}(b) where $\alpha =0.4$. Now the double-well
potential has been converted to a single well (see Fig.~\ref{potential}), so
the circuit behaves as a flux qubit in the single-well regime. Compared to the flux qubits in Figs.~%
\ref{energy_50}(a) and \ref{energy_50}(c), the energy levels in Figs.~\ref%
{energy_50}(b) and \ref{energy_50}(d) are less sensitive to the external
flux $f_{e}$, so the obtained flux qubits in the single-well regime are more robust against the flux
noise. However, because the smallest Josephson junction in the loop is
further reduced, the charge noise may become important~\cite{You-07}. To
suppress this charge noise, one can shunt a large capacitance to the
smallest junction to improve the quantum coherence of the qubit~\cite%
{yan-15,You-07,Steffen-10}.

Furthermore, let us consider the four-junction circuit with two identical smaller
Josephson junctions. In Fig.~\ref{energy_50}(e) where $\alpha=\beta =0.6$,
the lowest two levels are also well separated from other levels, but the
third level is not so separated from higher levels. Thus, from the
energy-level point of view, this four-junction circuit can be better used as
a flux qubit than a three-level system. In Fig.~\ref{energy_50}(f) where $%
\alpha =\beta =0.3$, the lowest three levels are well separated from other
levels. It seems that the four-junction circuit can be better used as a
three-level system. However, our calculations on transition matrix elements
indicate that the circuit can still be better used as a qubit, because only
the transition matrix element between the ground and first excited states is appreciably
large (see the next section).

In addition, we further consider the case of two different smaller Josephson junctions (i.e., $\alpha\neq\beta$) in the four-junction circuit. In the double-well regime [see Fig.~\ref{energy_50}(g), where $\alpha=0.5$ and $\beta=0.6$], the energy levels look similar to those in Fig.~\ref{energy_50}(e) and the lowest two levels can still be used as a qubit. Also, this qubit is less sensitive to the influence of the external magnetic field around the degeneracy point, because the energy levels are more flat than those in Fig.~\ref{energy_50}(e). In the single-well regime [see Fig.~\ref{energy_50}(h), where $\alpha=0.2$ and $\beta=0.3$], the lowest three levels are well separated from the higher levels. Moreover, in addition to the transition matrix element between the ground and first excited states, the transition matrix element between the first and second excited states is also larger (see the section below). Therefore, in the single-well regime, the four-junction circuit in the case of $\alpha\neq\beta$ can be better used as a quantum three-level system. This is different from the cases in Figs.~\ref{energy_50}(b), \ref{energy_50}(d) and \ref{energy_50}(f).

\vspace{0.3cm}\noindent \textbf{Transition matrix elements.} Now we consider
the time-dependent case with $f_{\mathrm{tot}}(t)\simeq f_{e}+f_{a}(t)$,
i.e., in addition to a static flux $f_{e}$, a time-dependent flux $%
f_{a}(t)\equiv \Phi _{a}(t)/\Phi _{0}$ is also applied to the four-junction
loop. In this case, $\xi \simeq f_{a}(t)$ when ignoring the very small
inductance-induced flux. For a small enough time-dependent flux, only the
first-order perturbation due to $\xi $ needs to be considered in Eq.~(\ref%
{Hfull}). Then, the Hamiltonian of the four-junction circuit in Eq.~(\ref%
{Hfull}) can be expressed as
\begin{equation}
H=H_{0}+H^{\prime }(t),
\end{equation}%
with $H_{0}$ given in Eq.~(\ref{H0}) and
\begin{eqnarray}
H^{\prime }(t)\! &\!=\!&\!f_{a}(t)E_{J}\bigg[\alpha b\sin \left( \frac{%
\varphi }{\sqrt{2}}+b_{+}\varphi _{+}+b_{-}\varphi _{-}\right)  \notag \\
&&+\alpha b\sin \left( -\frac{\varphi }{\sqrt{2}}+b_{+}\varphi
_{+}+b_{-}\varphi _{-}\right)  \notag \\
&&\!-\alpha b\sin \left( \frac{2\beta b_{+}}{\alpha +\beta -\lambda _{+}}%
\varphi _{+}+\frac{2\beta b_{-}}{\alpha +\beta -\lambda _{-}}\varphi
_{-}\right)  \notag \\
&&\!+\left( 2\alpha b+b+2\pi \right) \beta \sin \bigg(\frac{2\left( \alpha
-\lambda _{+}\right) b_{+}}{\alpha +\beta -\lambda _{+}}\varphi _{+}  \notag
\\
&&~~~~~~~~~+\frac{2\left( \alpha -\lambda _{-}\right) b_{-}}{\alpha +\beta
-\lambda _{-}}\varphi _{-}+2\pi f_{e}\bigg)\bigg].
\end{eqnarray}%
The time-dependent perturbation $H^{\prime }(t)$ can be rewritten as
\begin{equation}
H^{\prime }(t)=-I\Phi _{a}(t),
\end{equation}%
where
\begin{eqnarray}
I\! &\!=\!&\!-\frac{E_{J}}{\Phi _{0}}\bigg[\alpha b\sin \left( \frac{\varphi
}{\sqrt{2}}+b_{+}\varphi _{+}+b_{-}\varphi _{-}\right)  \notag \\
&&\!+\alpha b\sin \left( -\frac{\varphi }{\sqrt{2}}+b_{+}\varphi
_{+}+b_{-}\varphi _{-}\right)  \notag \\
&&\!-\alpha b\sin \left( \frac{2\beta b_{+}}{\alpha +\beta -\lambda _{+}}%
\varphi _{+}+\frac{2\beta b_{-}}{\alpha +\beta -\lambda _{-}}\varphi
_{-}\right)  \notag \\
&&\!+\left( 2\alpha b+b+2\pi \right) \beta \sin \bigg(\frac{2\left( \alpha
-\lambda _{+}\right) b_{+}}{\alpha +\beta -\lambda _{+}}\varphi _{+}  \notag
\\
&&~~~~~~~~~+\frac{2\left( \alpha -\lambda _{-}\right) b_{-}}{\alpha +\beta
-\lambda _{-}}\varphi _{-}+2\pi f_{e}\bigg)\bigg]
\end{eqnarray}%
is the current in the superconducting loop. Because
\begin{eqnarray}
(2\alpha b+b+2\pi )\beta \! &\!=\!&\!-\alpha b  \notag \\
\! &=&-\frac{2\pi \alpha \beta }{\alpha +\beta +2\alpha \beta },
\end{eqnarray}%
we can express the current $I$ as
\begin{eqnarray}
I\! &\!=\!&\!\frac{\alpha \beta }{\alpha +\beta +2\alpha \beta }\left( \frac{%
2\pi E_{J}}{\Phi _{0}}\right) [\sin \varphi _{1}+\sin \varphi _{2}+\sin
\varphi _{3}  \notag \\
&&\!~~~~~~~~~~~~~~~~~~~-\sin (\varphi _{1}+\varphi _{2}+\varphi _{3}+2\pi
f_{e})]  \notag \\
&=&\!\frac{1}{\alpha +\beta +2\alpha \beta }(\alpha \beta I_{1}+\alpha \beta
I_{2}+\beta I_{3}+\alpha I_{4}),
\end{eqnarray}%
where $I_{i}=I_{c}\sin \varphi _{1}$, with $i=1,2$ and $I_{c}=2\pi
E_{J}/\Phi _{0}$, $I_{3}=\alpha I_{c}\sin \varphi _{3}$, and $I_{4}=\beta
I_{c}\sin \varphi _{4}$ are Josephson supercurrents through the four
junctions. The phase drops $\varphi _{i}$ ($i=1,2,3$) are related to $%
\varphi $ and $\varphi _{\pm }$ by Eq.~(\ref{19}), and $\varphi _{4}$ is
constraint by the fluxoid quantization condition in the static-flux case,
i.e., $\varphi _{1}+\varphi _{2}+\varphi _{3}+\varphi _{4}+2\pi f_{e}=0$.

\begin{figure*}[tbp]
\includegraphics[scale=0.9]{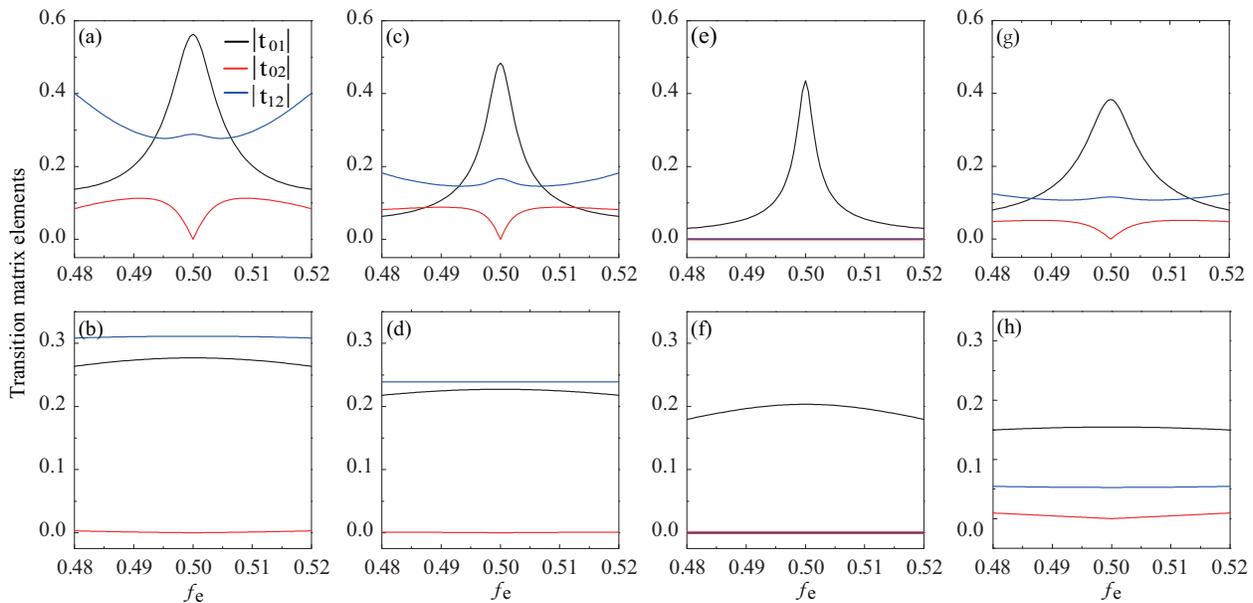}
\caption{(color online) Transition matrix elements $|t_{01}|$, $|t_{02}|$
and $|t_{12}|$ of the superconducting circuits (in units of $%
I_{c}\Phi_a^{(0)} $) versus the reduced static flux $f_{e}$. (a)~$\protect%
\alpha =0.7$ and (b)~$0.4$ in the case of three-junction circuit; (c)~$%
\protect\alpha =1$ and $\protect\beta =0.6$, (d)~$\protect\alpha =1$ and $%
\protect\beta =0.3$, (e)~$\protect\alpha=\protect\beta =0.6$, (f)~$%
\protect\alpha =\protect\beta =0.3$, (g)~$\protect\alpha =0.5$ and $\protect\beta %
=0.6$, and (h)~$\protect\alpha =0.2$ and $\protect\beta %
=0.3$ in the case of four-junction circuit.}
\label{transition_50}
\end{figure*}

Here we consider a microwave field with frequency $\omega _{a}$ applied to
the superconducting loop. The time-dependent magnetic flux in the loop can
be written as $\Phi _{a}(t)=\Phi _{a}^{(0)}\cos \omega _{a}t$. Then, with
the current $I$ available, the magnetic-dipole transition matrix elements
are calculated by%
\begin{equation}
t_{ij}=\langle i|I\Phi _{a}^{(0)}|j\rangle ,  \label{trans}
\end{equation}%
where $|i\rangle $ and $|j\rangle $ are eigenstates of the Hamiltonian $%
H_{0} $ in Eq.~(\ref{H0}).

Figure~\ref{transition_50} shows the transition matrix elements $|t_{01}|$, $%
|t_{02}|$, and $|t_{12}|$ of the three- and four-junction circuits as a
function of the reduced static flux $f_{e}$, where the subscripts $0$, $1$
and $2$ correspond to the ground state $|0\rangle $, the first excited state
$|1\rangle $, and the second excited state $|2\rangle $ of the system,
respectively. Similar to the three-junction circuit in Fig.~\ref%
{transition_50}(a) where $\alpha =0.7$, the four-junction circuit with $%
\alpha =1$ and $\beta =0.6$ (i.e., there is only one smaller Josephson junction in the circuit)
behaves as a ladder-type (namely, $\Xi $-type~\cite{Scully-05new})
three-level system at $f_{e}=1/2$, and a cyclic-type ($%
\Delta $-type~\cite{Liu-05}) three-level system at $f_{e}\neq 1/2$ [see Fig.~%
\ref{transition_50}(c)]. For the $\Xi $-type three-level system achieved when
$f_{e}=1/2$, the transition between the ground state $|0\rangle $ and the
second excited state $|2\rangle $ is not allowed, which is analogous to a
natural atom. However, for the $\Delta $-type
three-level system at $f_{e}\neq 1/2$, all transitions among $|0\rangle $, $%
|1\rangle $ and $|2\rangle $ are allowed. This is different from a natural
atomic system~\cite{Liu-05}. When the smallest Josephson junction is further
reduced, $|t_{02}|$ is greatly suppressed. Now both three- and four-junction
circuits behave more like a $\Xi $-type three-level system in the whole
region of $f_{e}$ shown in Figs.~\ref{transition_50}(b) and \ref%
{transition_50}(d).

As for the four-junction circuit with two identical smaller Josephson junctions ($\alpha=\beta$), while
$|t_{01}|$ remains appreciably large, the transition between $|0\rangle$ and
$|2\rangle$ as well as the transition between $|1\rangle$ and $|2\rangle$
are greatly reduced (i.e., $|t_{02}|\approx 0$ and $|t_{12}|\approx 0$) in
the whole region of $f_e$ shown in Figs.~\ref{transition_50}(e) and \ref%
{transition_50}(f). Now, in either double- or single-well regime,
the four-junction circuit can be well used as a qubit, because the state
leakage from the qubit subspace to the third level is suppressed. This is an apparent
advantage of the four-junction circuit over the three-junction circuit when used as a qubit.

When the two smaller Josephson junctions in the four-junction circuit become different
(i.e., $\alpha\neq\beta$), in addition to $|t_{01}|$, both $|t_{02}|$ and $|t_{12}|$ become nonzero except for the degeneracy point [see Figs.~\ref{transition_50}(g) and \ref{transition_50}(h)]. This circuit behaves very different from the circuit with two identical smaller junctions
[comparing Fig.~\ref{transition_50}(g) with Fig.~\ref{transition_50}(e), and comparing Fig.~\ref{transition_50}(h) with
Fig.~\ref{transition_50}(f)], but it is similar to the three-junction circuit and the four-junction circuit with only one smaller junction [comparing Fig.~\ref{transition_50}(g) with Figs.~\ref{transition_50}(a) and \ref{transition_50}(c), and comparing Fig.~\ref{transition_50}(h) with Figs.~\ref{transition_50}(b) and \ref{transition_50}(d)]. However,
when the distribution of the energy levels is also taken into account (see Fig.~\ref{energy_50}), the four-junction circuit with $\alpha\neq\beta$ can be better used as a quantum three-level system (qutrit) in the single-well regime.  This is very different from the three-junction circuit and the four-junction circuit with only one smaller junction, which can be better used as a qubit in the single-well regime. Therefore, as compared to the three-junction circuit, the four-junction circuit can provide more choices to achieve different quantum systems.

\vspace{1cm}\noindent \textbf{\large Summary}

\vspace{.1cm}\noindent We have developed a theory for the four-junction
superconducting loop pierced by an externally applied magnetic flux. When
the loop inductance is considered, the derived Hamiltonian of this
four-junction circuit can be written as the sum of two parts, one of which
is the Hamiltonian of a harmonic oscillator with a very large frequency.
This makes it feasible to employ the adiabatic approximation to eliminate
the degree of freedom of the harmonic oscillator in the total Hamiltonian.
Also, this theory can be used to study the case when the applied magnetic-flux bias becomes time-dependent.
In the case of static flux bias, the total Hamiltonian of the four-junction
circuit is reduced to the Hamiltonian of the superconducting qubit. When the
flux bias is time-dependent, the total Hamiltonian of the four-junction
circuit can be reduced to the Hamiltonian of the superconducting qubit plus
a perturbation related to the applied time-dependent flux. Then, we can
calculate the energy spectrum and the transition matrix elements of the
four-junction superconducting circuit.

In conclusion, we have studied the four-junction superconducting circuit in
both double- and single-well regimes. In addition to the similarity with the
three-junction circuit, we show the difference of the four-junction circuit
from its three-junction analogue. Also, we demonstrate its
advantages over the three-junction circuit.
Owing to the one additional Josephson junction in the circuit, the physical properties of the four-junction circuit become richer than those of the three-junction circuit. For instance, in the case of four-junction circuit with only one smaller Josephson junction, the circuit has a broader parameter range to achieve a flux qubit in the double-well regime than the three-junction circuit does. Moreover, in the case of four-junction circuit with two identical smaller junctions, the circuit can be used as a qubit better than the three-junction circuit in both double- and single-well regimes. This is because among the lowest three eigenstates of the four-junction circuit, only the transition matrix element between the ground and first excited states is appreciably large, while other two elements become zero.
These properties of the four-junction circuit can suppress the state leakage from the qubit subspace to the second excited state, and the circuit with these parameters is thus expected to have better quantum coherence when used as a qubit.

\vspace{1cm}\noindent \textbf{\large Methods}

\vspace{.1cm}\noindent \textbf{Three-junction circuit with a time-dependent
magnetic flux.} To compare with our four-junction results, we also consider
a three-junction superconducting loop pierced by a time-dependent total
magnetic flux $\Phi_{\mathrm{tot}}(t)$ [see Fig.~\ref{system}(b)], because
no explicit derivation exists in the literature for this time-dependent
case. The directions of the phase drops $\varphi _{i}$ ($i=1,2,3$) through
the three Josephson junctions are chosen as in Ref.~\cite{Orlando-99}, which
are constrained by the following fluxoid quantization condition:
\begin{equation}
\varphi _{1}-\varphi _{2}+\varphi _{3}+2\pi f_{\mathrm{tot}}(t)=0,
\label{A1}
\end{equation}
where $f_{\mathrm{tot}}(t)=\Phi _{\mathrm{tot}}(t)/\Phi _{0}$. Here we
assume that two larger junctions have identical capacitance $C$ and coupling
energy $E_{J}$, while the smaller junction has capacitance $\alpha C$ and
coupling energy $\alpha E_J$, with $0<\alpha<1$.

Similar to the four-junction circuit, we introduce a phase transformation
\begin{eqnarray}
\varphi _{p}\! &\!=\!&\!\frac{\varphi _{1}+\varphi _{2}}{2},  \notag \\
\ \varphi _{m}\! &\!=\!&\!\frac{\varphi _{1}-\varphi _{2}}{2}+\frac{2\pi
\alpha }{1+2\alpha }\xi ,  \label{PT-1}
\end{eqnarray}%
where $\xi \equiv f_{\mathrm{tot}}(t)-f_{e}$, with $f_{e}=\Phi _{e}/\Phi
_{0} $ being the reduced static magnetic flux applied to the superconducting
loop. The Hamiltonian of the three-junction circuit can be derived as
\begin{equation}
H=\!2E_{C}P_{p}^{2}+\frac{2E_{C}}{(1+2\alpha )}P_{m}^{2}+U(\varphi
_{p},\varphi _{m},\xi )+H_{\mathrm{osc}},
\end{equation}%
where $E_{C}=e^{2}/(2C)$,
\begin{eqnarray}
U\! &\!=\!&\!E_{J}\bigg\{2+\alpha -2\cos \varphi _{p}\cos \left( \ \varphi
_{m}-\frac{2\pi \alpha }{1+2\alpha }\xi \right)  \notag \\
&&~~~~ -\alpha \cos \left( 2\varphi _{m}+\frac{2\pi }{1+2\alpha }\xi +2\pi
f_{e}\right) \bigg\},
\end{eqnarray}
and
\begin{equation}
H_{\mathrm{osc}}=\frac{E_{C}(1+2\alpha )}{\pi ^{2}\alpha }P_{\xi }^{2}+\frac{%
\Phi _{0}^{2}}{2L}(\xi -f_{a})^{2}.  \label{osc}
\end{equation}%
Quantum mechanically, the canonical momenta can be
written as $P_{p}=-i\hbar \partial /\partial \varphi _{p}$, $P_{m}=-i\hbar
\partial /\partial \varphi _{m}$, and $P_{\xi }=-i\hbar \partial /\partial
\varphi _{\xi }$ in the canonical-coordinate representation.

The angular frequency of the harmonic oscillator given in Eq.~(\ref{osc}) is
\begin{equation}
\omega _{\mathrm{osc}}=\sqrt{\frac{1+2\alpha }{\alpha CL}}.
\end{equation}%
Using the parameters achieved in experiments~\cite{Bertet-05,Zant-94}, we
have $\alpha \sim 0.7$, $C\sim 8~\mathrm{fF}$, and $L\sim 10~\mathrm{pH}$,
so one has $\omega _{\mathrm{osc}}/2\pi \sim 10^{3}$~GHz, which is much
larger than the energy gap $\Delta \sim 1$-$10$~GHz of the three-junction flux
qubit (see, e.g., Ref.~\cite{Chiorescu-03}). If the time-dependent magnetic
flux is the usually applied microwave field, the oscillator can indeed be
regarded as being in the ground state at a low temperature, as analyzed for
the four-junction flux qubit in the main text. Then, the Hamiltonian of the
three-junction circuit can be reduced to
\begin{equation}
H=2E_{C}P_{p}^{2}+\frac{2E_{C}}{(1+2\alpha )}P_{m}^{2}+U(\varphi
_{p},\varphi _{m},\xi ).  \label{A9}
\end{equation}

Because $L$ is small in a three-junction flux qubit~\cite{Orlando-99}, we
can ignore the flux generated by the loop inductance. Thus, when only a
static flux is applied to the loop, $f_{\mathrm{tot}}(t)\simeq f_{e}$, i.e.,
$\xi \simeq 0$. The phase transformation in Eq.~(\ref{PT-1}) becomes
\begin{equation}
\varphi _{p}=\frac{\varphi _{1}+\varphi _{2}}{2},~~~\varphi _{m}=\frac{%
\varphi _{1}-\varphi _{2}}{2},  \label{A10}
\end{equation}%
and the Hamiltonian of the circuit in Eq.~(\ref{A9}) is reduced to
\begin{eqnarray}
\ H_{0}\! &\!=\!&\!2E_{C}P_{p}^{2}+\frac{2E_{C}}{(1+2\alpha )}%
P_{m}^{2}+E_{J}[2+\alpha  \label{A11} \\
&&-2\cos \varphi _{p}\cos \varphi _{m}-\alpha \cos (2\varphi _{m}+2\pi
f_{e})],~~~~  \notag
\end{eqnarray}%
which is the Hamiltonian of the three-junction flux qubit derived in Ref.~%
\cite{Orlando-99}.

For the time-dependent case with $f_{\mathrm{tot}}(t)\simeq f_{e}+f_{a}(t)$,
$\xi \simeq f_{a}(t)$, where $f_{a}(t)\equiv \Phi _{a}(t)/\Phi _{0}$ is the
reduced time-dependent magnetic flux applied to the three-junction loop.
When the time-dependent magnetic flux is small enough, only the first-order
perturbation due to $\xi $ needs to be considered, and the Hamiltonian of
the circuit in Eq.~(\ref{A9}) can be expressed as
\begin{equation}
H=H_{0}+H^{\prime }(t),
\end{equation}%
with $H_{0}$ given in Eq.~(\ref{A11}) and $H^{\prime }(t)=-I\Phi _{a}(t)$,
where
\begin{equation}
I=\frac{2\pi \alpha }{1+2\alpha }\frac{E_{J}}{\Phi _{0}}\left[ 2\cos \varphi
_{p}\sin \varphi _{m}-\sin \left( 2\varphi _{m}+2\pi f_{e}\right) \right]
\end{equation}%
is the current in the three-junction loop~\cite{Liu-14}. Using Eq.~(\ref{A10}%
) and the fluxoid quantization condition in the static-flux case (i.e., $%
\varphi _{1}-\varphi _{2}+\varphi _{3}+2\pi f_{e}=0$), the current $I$ can
also be rewritten as
\begin{eqnarray}
I\! &\!=\!&\!\frac{\alpha }{1+2\alpha }\left( \frac{2\pi E_{J}}{\Phi _{0}}%
\right) [\sin \varphi _{1}-\sin \varphi _{2}  \notag \\
&&\!~~~~~~~~~~~~~~~~~~-\sin \left( \varphi _{1}-\varphi _{2}+2\pi
f_{e}\right) ]  \notag \\
&=&\!\frac{1}{1+2\alpha }(\alpha I_{1}-\alpha I_{2}+I_{3}),
\end{eqnarray}%
where $I_{i}$ is the Josephson supercurrent through each junction. Moreover,
as in Eq.~(\ref{trans}), the magnetic-dipole transition matrix elements are
calculated by $t_{ij}=\langle i|I\Phi _{a}^{(0)}|j\rangle $, where $%
|i\rangle $ and $|j\rangle $ are eigenstates of the Hamiltonian $H_{0}$ in
Eq.~(\ref{A11}).

\vspace{1cm}\noindent \textbf{Acknowledgement}

\noindent This work is supported by the NSAF Grant No. U1330201, the
National Natural Science Foundation of China Grant No. 91121015, and the
National Basic Research Program of China Grant No. 2014CB921401.





\end{document}